\documentclass[twocolumn,aps,amsmath,amssymb,floatfix,superscriptaddress,showpacs]{revtex4}
\usepackage{graphicx}
\begin{document}
\title{Log-Networks}
\author{P. L. Krapivsky}
\email{paulk@bu.edu}
\affiliation{Center for Polymer Studies and  
Department of Physics, Boston University, Boston, MA 02215}
\author{S.~Redner}
\email{redner@bu.edu}\altaffiliation{Permanent address: 
Department of Physics, Boston University, Boston, MA 02215}
\affiliation{Theory Division  and Center for Nonlinear Studies, 
Los Alamos National Laboratory, Los Alamos, New Mexico 87545}
\begin{abstract}
  We introduce a growing network model in which a new node attaches to a
  randomly-selected node, as well as to all ancestors of the target node.
  This mechanism produces a sparse, ultra-small network where the average node
  degree grows logarithmically with network size while the network diameter
  equals 2.  We determine basic geometrical network properties, such as the
  size dependence of the number of links and the in- and out-degree
  distributions.  We also compare our predictions with real networks where
  the node degree also grows slowly with time --- the Internet and the
  citation network of all Physical Review papers.

\end{abstract}


\maketitle

\section{Introduction} 

Many networks in nature and technology are sparse, {\it i.e.}, the average
node degree is much smaller than the total number of nodes $N$
\cite{rev,book}.  Widely studied classes of networks, such as regular grids,
random graphs, and scale-free networks, are maximally sparse, as the average
node degree remains finite as $N\to\infty$.  However, in other examples of
real sparse networks, such as the Internet, the average node degree grows,
albeit very slowly, with system size.  Motivated by this observation, we
introduce a simple network growth mechanism of copying that naturally
generates sparse networks in which the average node degree diverges
logarithmically with system size.  We dub these log-networks.  Related
results appear in previous investigations of related models
\cite{alex,grandson}, while models where a slowly increasing ratio of links
to nodes is imposed externally have also been considered \cite{DM,sen}.

To motivate the copying mechanism for log-networks let us recall the growing
network with redirection (GNR) \cite{KR01}.  The GNR is built by adding nodes
according to the following simple rule.  Each new node initially selects an
earlier ``target'' node at random.  With a specified probability a link from
the new node to the target node is created; with a complementary probability,
the link is re-directed to the ancestor node of the target.  Although the
target node is chosen randomly, the redirection mechanism generates an
effective preferential attachment because a high-degree node is more likely
to be the ancestor of a randomly-selected node.  By this feature, redirection
leads to a power-law degree distribution for the network.  The GNR thus
provides an appealingly simple mechanism for preferential attachment, as well
as an extremely efficient way to simulate large scale-free networks
\cite{KR021}.

The growing network with redirection is a simplification of a previous model
\cite{kum} which was proposed to mimic the copying of links in the world-wide
web.  In this web model, a new node links to a randomly-chosen target node
and also to its ancestor nodes (subject to a constraint on the maximum number
of links created).  In the context of citations, copying is (regrettably)
even more natural, as it is easier merely to copy the references of a cited
paper, rather than to look at the original references \cite{sim}.  As the
literature grows, the copying mechanism will necessarily lead to later
publications having more references than earlier publications.

In the following sections, we analyze a growing network model with copying
(GNC).  We consider a model with no global bound on the number of links
emanating from a new node.  We shall see that this simple copying mechanism
generates log-networks.  We will use the master equation approach to derive
basic geometric properties of the network.  We then compare our prediction
about the logarithmic growth of the average degree with data from Physical
Review citations.

\section{GNC Model}

\begin{figure}[ht] 
 \vspace*{0.cm}
 \includegraphics*[width=0.24\textwidth]{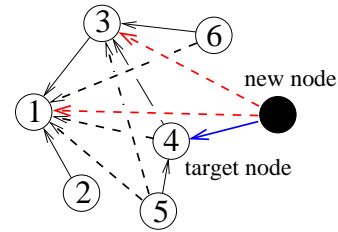}
 \caption{Illustration of the growing network with copying (GNC).  The time
   order of the nodes is indicated.  Initial links are solid and links to
   ancestor nodes are dashed.  Later links partially obscure earlier links.
   The new node initially attaches to random target node 4, as well as to its
   ancestors, 1 and 3.}
\label{C}
\end{figure}

We now define the GNC model precisely.  The network grows by adding nodes one
at a time.  A newly-introduced node randomly selects a target node and links
to it, as well as to all ancestor nodes of the target node (Fig.~\ref{C}).

If the target node is the initial root node, no additional links are
generated by the copying mechanism.  If the newly-introduced node were to
always choose the root node as the target, a star graph would be generated.
On the other hand, if the target node is always the most recent one in the
network, all previous nodes are ancestors of the target and the copying
mechanism would give a complete graph.  Correspondingly, the total number of
links $L_N$ in a network of $N$ nodes can range from $N-1$ (star graph) to
$N(N-1)/2$ (complete graph).  Notice also that the number of outgoing links
from each new node (the out-degree) can range between 1 and the current
number of nodes.

\section{Network Structure} 

We now study geometric properties of the GNC model by the master equation
approach.  We determine how the total number of links $L$ grows with $N$, as
well as the in-degree, out-degree, and the joint in/out-degree distributions.

\subsection{Total Number of Links}

Let $L(N)$ be the average value of the total number of links in a network of
$N$ nodes.  If a newly-introduced node selects a target node with $j$
ancestors, then the number of links added to the network will be $1+j$.
Therefore the average total number of links satisfies
\begin{eqnarray}
\label{LN}
L(N+1)&=&L(N)+\frac{1}{N}\Big\langle \sum_{\alpha}(1+j_\alpha)\Big\rangle\nonumber\\
      &=&L(N)+1+\frac{L(N)}{N}.
\end{eqnarray}
The factor $N^{-1}$ in the first line assures that a target node $\alpha$ is
selected uniformly from among all $N$ nodes, and we obtain the second line by
employing the sum rule $\langle\sum_{\alpha}j_\alpha\rangle=L$.

Dividing Eq.~(\ref{LN}) by $N+1$ gives
\begin{equation*}
\frac{L(N+1)}{N+1}-\frac{L(N)}{N}=\frac{1}{N+1},
\end{equation*}
and then summing both sides from 1 to $N-1$ gives the solution
\begin{equation}
\label{LN-sol}
L(N)=N\left(H_N-1\right)
\end{equation}
Here $H_N=\sum_{n=1}^{N}n^{-1}$ is the harmonic number.  (For concreteness,
we assume that the network starts with a single node, so that $L(1)=0$.)~
Using the asymptotics of the harmonic numbers \cite{knuth} we find
\begin{equation}
\label{LN-sol-asymp}
L(N)=N\ln N-N(1-\gamma)+\frac{1}{2}-\frac{1}{12 N}+\ldots\,,
\end{equation}
where $\gamma=0.57721566\ldots$ is the Euler constant.  The leading asymptotic
behavior of $N\ln N$ can also be obtained more easily from Eq.~(\ref{LN}) by
taking the continuum approximation and solving the resulting differential
equation.

\begin{figure}[ht] 
 \vspace*{0.cm}
 \includegraphics*[width=0.45\textwidth]{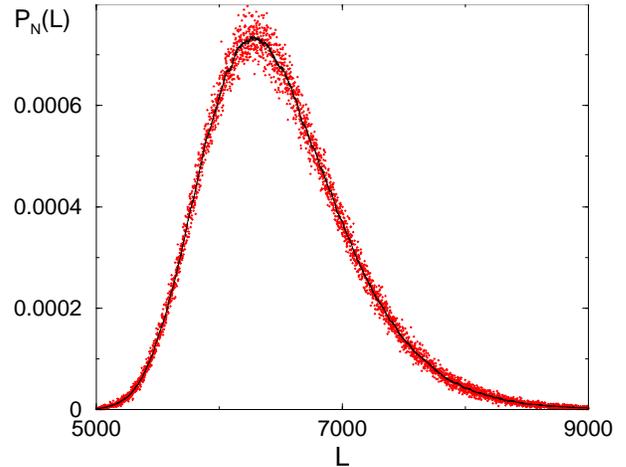}
 \caption{Distribution of the number of links $P_N(L)$ for $10^5$
 realizations of a GNC of $N=1000$ sites.  Shown are both the raw data and
 the result of averaging the data over a 1\% range.  For this value of $N$,
 the mean number of links is 6485.56, while Eq.~(\ref{LN-sol-asymp}) gives
 6485.47\ldots. }
\label{PL}
\end{figure}

Thus we conclude that the average degree of the network grows logarithmically
with the system size; that is, the copying mechanism generates a log-network.
This simple phenomenon is one of our major results.

We now briefly discuss the probability distribution of the total number of
links $P_N(L)$ for a network of $N$ nodes.  Simulations show that the
distribution is asymmetric and quite broad (Fig.~\ref{PL}).  To understand
the origin of the asymmetry, notice that both the extreme cases of the star
graph ($L=N-1$) and the complete graph ($L=N(N-1)/2$) each occur with
probability
\begin{eqnarray*}
P_N(L)=\frac{1}{(N-1)!}\,,
\end{eqnarray*}
because each new node must select one specific target node.  Therefore the
distribution of the total number of links vanishes much more sharply near the
lower cutoff.

Near the peak however, the distribution $P_N(L)$ is symmetric about the
average.  More precisely, when the deviation from the average $L(N)=\sum_L
L\,P_N(L)$ is of the order of $\Sigma(N)=\sqrt{\sum_L [L-L(N)]^2\,P_N(L)}$,
the distribution approaches a symmetric Gaussian shape.  The value of the
standard deviation as $N\to\infty$ is
\begin{equation}
\label{SN-sol-asymp}
\Sigma(N)\to C\,N, \qquad C=\sqrt{2-\pi^2/6}=0.595874\ldots\,,
\end{equation}
as derived in Appendix \ref{ap}.  The relative width of the distribution is
measured by the standard deviation $\Sigma(N)$ divided by the average $L(N)$;
this ratio approaches zero as $(\ln N)^{-1}$ when $N\to\infty$, so that
fluctuations die out slowly.  This slow decay of fluctuations explains why
the distribution (Fig.~\ref{PL}) remains wide for large $N$ and why it looks
asymmetric near the peak.

The GNC model can be extended to allow for wider copying variability.  For
example, instead of linking to one initial random target node, we can link to
$m$ random initial targets.  Further, we can link to each target node with
probability $p$ and to each of the corresponding ancestors with probability
$q$. For this general $(m,p,q)$ model, the analog of Eq.~(\ref{LN}) is
\cite{note} 
\begin{equation}
\label{LN-mpq}
L(N+1)=L(N)+\frac{m}{N}\Big\langle\sum_{\alpha}(p+qj_\alpha)\Big\rangle
\end{equation}
which reduces to $\frac{dL}{dN}=mp+mq\frac{L}{N}$ in the continuum
approximation. The asymptotic growth of the average total number of links
crucially depends on the parameter $mq$:
\begin{equation}
\label{LN-mpq-sol}
L(N)= 
\begin{cases}
{\displaystyle \frac{mp}{1-mq}\,N}  & {\rm for} \quad mq<1;\cr
{\displaystyle mp\,N \ln N}         & {\rm for} \quad mq=1;\cr
{\displaystyle \propto N^{mq}}      & {\rm for} \quad  mq>1.
\end{cases} 
\end{equation}
Thus incomplete copying leads to an average node degree that is independent
of $N$ when $mq<1$, while marginal logarithmic dependence is recovered when
$mq=1$.  There is also a pathology for $mq>2$, as the number of links in the
network would exceed that of a complete graph with the same number of nodes.
In this case, it is not possible to accommodate all the links specified by
the copying rule without having a multigraph, {\it i.e.}, allowing for more
than one link between a given pair of nodes.

\subsection{Comparison with Empirical Data}

\begin{figure}[ht] 
 \vspace*{0.cm}
 \includegraphics*[width=0.45\textwidth]{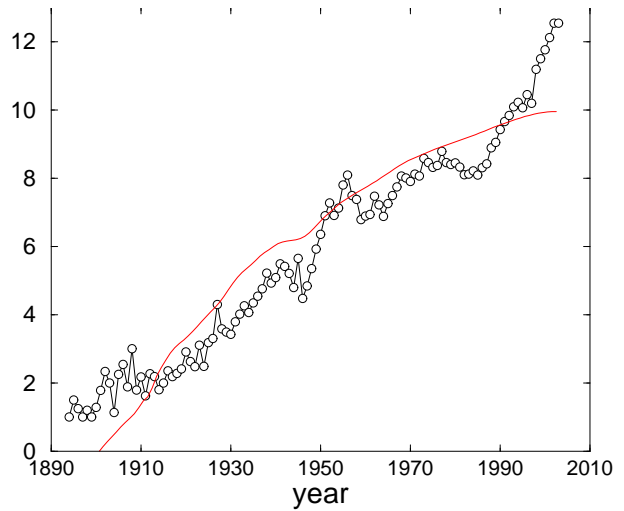}
 \caption{Average number of references in the reference list of Physical
   Review papers published in each year ($\circ$).  Also shown as a smooth
   curve is the logarithm of the cumulative number of Physical Review papers
   that were published up to each year.  The value 5 is subtracted from the
   latter data to make the two datasets lie in the same range.}
\label{ref-compare}
\end{figure}

We now present empirical data from the citation network of Physical Review to
test whether the average degree of these networks grows with time, and if so,
whether the growth is consistent with log-networks. Data from all issues of
Physical Review journals is available, encompassing a time span of 110 years
\cite{PR}.  From this data, we have the following evidence that citations may
be described as a log-network.  Specifically, the average number of
references in the reference list of each Physical Review paper grows
systematically with time and is consistent with a linear increase
(Fig.~\ref{ref-compare}).  Additionally, the number of Physical Review papers
published in a given year roughly grows exponentially with time \cite{PR}.
Thus the cumulative number of Physical Review papers up to a given year also
grows exponentially.  As a result, the number of references should grow
logarithmically with the total number of available papers.  This behavior is
reasonably consistent with the data of Fig.~\ref{ref-compare}.

In a related vein, the average number of coauthors per paper has grown slowly
with time, due in part, to the growing trend for collaborative research and
the continuing ease of long-distance scientific interaction.  While
co-authorship and other collaboration networks have recently been
investigated (see, {\it e.g.}, \cite{new,dor,LJL}), the analysis has
primarily been on network properties at a fixed time.  There is, however, one
study of the number of mathematics papers with 1, 2, and more authors since
1940 \cite{ENP}.  This data shows that the fraction of singly-authored papers
is decreasing systematically, while the number of multiple-authored papers is
steadily growing.  Thus it should be interesting to track the time dependence
of the number of co-authors in scientific publications from the current
studies of collaboration networks.

Interestingly, the Internet and the world wide web exhibit certain
similarities with log-networks.  For example, the total number of links
exceeds the total number of nodes in the world wide web by about an order of
magnitude \cite{broder,don}.  Similarly for the Internet, specifically for
the Autonomous Systems (AS) graph, the average number of links per node is
also growing slowly but systematically with time \cite{BNC}.  Qualitatively
these behaviors are consistent with our expectations from log-networks.  It
is not still possible to reach definitive conclusions about the precise
growth rate on $N$ since the available data for the AS graph \cite{BNC}
covers a time period when the total number of ASes has increased only by a
factor of 4 (from $N=3060$ in 1997 to $N=12155$ in 2001).

\subsection{In-degree distribution} 

\begin{figure}[ht] 
 \vspace*{0.cm}
\center{\includegraphics*[width=0.25\textwidth]{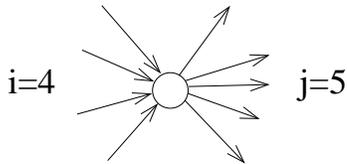}}
\caption{A node with in-degree (number of incoming links) $i=4$, out-degree
  (number of outgoing links) $j=5$, and total degree $k=9$.}
\label{degrees}
\end{figure}

By its very construction, the links of the GNC network are directed and thus
there is an in-degree $i$ and an out-degree $j$ for each node
(Fig.~\ref{degrees}), and thus two distinct corresponding degree
distributions.  In this subsection, we study the in-degree distribution.

Let $P_i(N)$ be the average number of nodes with in-degree $i$ in a network
consisting of $N$ total nodes.  This distribution satisfies 
\begin{equation}
\label{Pi}
P_i(N+1)=P_i(N)-\frac{i+1}{N}\,P_i(N)+\frac{i}{N}\,P_{i-1}(N)+\delta_{i,0}
\end{equation}
The loss term accounts for the following two processes: (a) either a node of
in-degree $i$, or (b) any of its $i$ daughter nodes was chosen as the target.
Either of these processes leads to the loss of a node with in-degree $i$.
The total loss rate of $P_i(N)$ is thus $(i+1)/N$.  The gain term is
explained similarly, and the last term on the right-hand side of
Eq.~(\ref{Pi}) describes the effect of the introduction of a new node with no
incoming links.  Finally, notice that Eqs.~(\ref{Pi}) hold for $i\leq N$.
When $i=N$, there is no longer a loss term and the master equation reduces to
$P_N(N+1)=P_{N-1}(N)=1$.  This accounts for the fact that the root node is
necessarily linked to all other nodes and therefore there is one node with
degree $N-1$ in a network of $N$ nodes.

We compute the in-degree distribution by induction.  Solving for the first
few $P_i(N)$ for small $i$ directly, we find a simple form for the general
case that we then check solves the master equation (\ref{Pi}).  We thus find
\begin{equation}
\label{pi}
P_i(N)=\frac{N}{(i+1)(i+2)} \qquad{\rm for}~ i<N-1,
\end{equation}
while $P_i(N)=0$ for $i\geq N$.

The asymptotic $i^{-2}$ decay agrees with the logarithmic divergence for the
average node degree from the previous section, and perhaps explains the
proliferation of exponent values close to 2 for the in-degree distribution
that are observed in empirical studies of collaboration networks
\cite{new,dor,LJL} and in the world wide web \cite{AH,BA,www,broder,don}.  In
particular, a comprehensive study by by Broder et al.~\cite{broder} reports
an exponent value 2.09, while a recent work by Donato et al.~\cite{don}
(relying on the WebBase project at Stanford \cite{stanford}) quotes an
exponent value 2.1.

\subsection{Out-degree distribution} 

\begin{figure}[ht] 
 \vspace*{0.cm}
 \includegraphics*[width=0.325\textwidth]{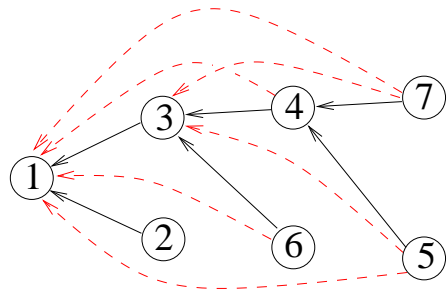}
 \caption{Genealogical tree representation of the network of Fig.~\ref{C},
   with nodes arranged in layers left to right according to their out-degree.
   The initial layer contains only the root node.  The number of nodes in
   subsequent layers increases as the network grows.  The initial links are
   shown as solid arrows and the copied links as dashed arrows.}
\label{out}
\end{figure}

To determine the out-degree distribution, it is helpful to think of the
network as a genealogical tree, as illustrated in Fig.~\ref{out} (see also
\cite{KR01} for this construction).  Initially the network consists of one
root node.  Subsequent nodes that attach to the root node will have
out-degree 1 and lie in the first layer.  Similarly, a new node that attaches
to a node with out-degree 1 lies in the second layer.  By the copying
mechanism, nodes in this second layer also link to the root and therefore
have out-degree 2.  Nodes in the $n^{\rm th}$ layer directly attach to a node
in the $(n-1)^{\rm st}$ layer and, by virtue of copying, also attach to one
node in every previous layer.  Thus $n^{\rm th}$ layer nodes have out-degree
$n$.  We now use this genealogical tree picture to determine the out-degree
distribution of the network.

Let $Q_j(N)$ be the average number of nodes with out-degree $j$ in a network
consisting of $N$ nodes.  By definition $Q_0(N)\equiv 1$.  On the other hand,
the number of nodes with out-degree $j\geq 1$ grows each time a node with
out-degree $j-1$ is selected as the target node.  The out-degree distribution
thus satisfies the master equation
\begin{equation}
\label{Qj}
Q_j(N+1)=Q_j(N)+\frac{1}{N}\,Q_{j-1}(N).
\end{equation}
This equation applies even for $j\!=\!0$ if we set \mbox{$Q_{-1}(N)\equiv
  0$}.  Using the recursive nature of these equations, we first solve for
$Q_1(N)$, then $Q_2(N)$, {\it etc}, and ultimately the out-degree
distribution for all $j$.  This procedure gives
\begin{equation}
\label{Qj-sol}
Q_j(N+1)=\sum_{1\leq m_1<\ldots<m_j\leq N}\frac{1}{m_1\times\ldots\times m_j}
\end{equation}
Equivalently, we can recast the $j$-fold sums into simple sums, although
the results look less neat.  For example,
\begin{equation*}
Q_2(N+1)=\frac{1}{2}\left[(H_N)^2-H_N^{(2)}\right]
\end{equation*}
where $H_N^{(2)}=\sum_{n=1}^{N}n^{-2}$.  The asymptotic behaviors of $H_N$,
$H_N^{(2)}$, and other generalized harmonic numbers are known \cite{knuth},
and the resulting asymptotics of the out-degree distribution are
\begin{eqnarray*}
Q_1(N+1)&=&H_N=\ln N+\gamma+\frac{1}{2N}-\frac{1}{12 N^2}+\ldots\\
Q_2(N+1)&=&\frac{1}{2}\,(\ln N)^2+\gamma\,\ln N
+\frac{1}{2}\left[\gamma^2-\frac{\pi^2}{6}\right]+\ldots
\end{eqnarray*}
and analogous results hold for $Q_j(N)$ for larger $j$.

If we merely want to establish the leading asymptotic behavior, we can
replace the summation in (\ref{Qj-sol}) by integration.  This then leads to
the simple result
\begin{equation}
\label{Qj-sol-cont}
Q_j(N)\to \frac{(\ln N)^j}{j!}.
\end{equation}
Alternatively, we can derive this result within a continuum approach by
replacing finite differences by derivatives in the large-$N$ limit of
Eq.~(\ref{Qj}).  The procedure recasts the discrete master equations into
the differential equations
\begin{equation*}
\frac{dQ_j}{dN}=\frac{1}{N}\,Q_{j-1}(N)
\end{equation*}
whose solution is indeed (\ref{Qj-sol-cont}).

The Poisson form of the out-degree distribution contradicts the commonly
presumed power-law form.  There is previous literature by Broder {\it et al.}
that suggested that the out-degree distribution has a power-law tail, with
exponent close to 2.7 \cite{broder}.  However, this work also noted that a
power-law is not a good fit to the data and that the out-degree distribution
may possibly follow a Poisson distribution.  In fact, the analysis of
Ref.~\cite{don} that is based on more recent data on the structure of the web
\cite{stanford} convincingly shows that a power law does not fit the
out-degree distribution.

\subsection{Joint degree distribution} 

We define the joint degree distribution $N_{i,j}(N)$ as the average number of
nodes with in-degree $i$ and out-degree $j$ in a network of $N$ nodes.  The
in- and out-degree distributions can then be distilled from the joint
distribution via $P_i(N)=\sum_{j}N_{i,j}(N)$ and $Q_j(N)=\sum_{i}N_{i,j}(N)$.
Furthermore, the average number of nodes with total degree $k$ is simply
given by $N_k(N)=\sum_{i+j=k}N_{i,j}(N)$.

\begin{figure}[ht] 
 \vspace*{0.cm}
\center{\includegraphics*[width=0.45\textwidth]{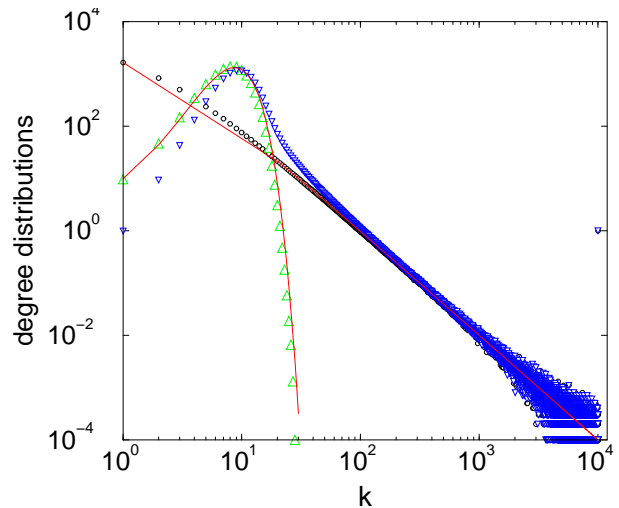}}
\caption{The in- ($\circ$), out- ($\bigtriangleup$), and total
 ($\bigtriangledown$) degree distributions for $10^4$ realizations of a
 network of $N=10^4$ nodes.  Notice that there is always one node with total
 degree equal to 1 and one node with in- and total degree equal to $N$.  The
 smooth curve that follows in in-degree data is the asymptotic prediction
 (\ref{pi}), while the curve that follows the out-degree data is the
 asymptotic prediction (\ref{Qj-sol-cont}).}
\label{deg-dist}
\end{figure}

The joint degree distribution satisfies
\begin{eqnarray}
\label{Nij}
N_{i,j}(N\!+\!1)
&\!\!=\!\!&N_{i,j}(N)
+\frac{i}{N}\,N_{i\!-\!1,j}(N)-\frac{i\!+\!1}{N}\,N_{i,j}(N)\nonumber\\
&+&\frac{1}{N}\,Q_{j\!-\!1}(N)\,\delta_{i,0}
\end{eqnarray}
which is an obvious generalization of the governing equations (\ref{Pi}) and
(\ref{Qj}) for the separate in- and out-degree distributions.  Because of the
presence of the last out-degree term on the right-hand side of equation
(\ref{Nij}), the scaling of the joint degree distribution with system size
does not hold -- $N_{i,j}(N)\ne N\,n_{i,j}$.  Therefore we cannot reduce
(\ref{Nij}) to an $N$-independent recursion.

Nevertheless, Eqs.~(\ref{Nij}) still have the important simplifying feature of
being recursive and thus soluble in an inductive fashion.  Thus, for example,
for $i=0$ we have
\begin{equation*}
N\,N_{0,j}(N+1)=(N-1)\,N_{0,j}(N)+Q_{j-1}(N)
\end{equation*}
from which 
\begin{equation}
\label{N0j}
N_{0,j}(N+1)=\frac{1}{N}\sum_{M=1}^NQ_{j-1}(M)
\end{equation}
For $i\geq 1$, we rewrite Eqs.~(\ref{Nij}) in the form
\begin{equation}
\label{Nij1}
N N_{i,j}(N+1)=(N-i-1)N_{i,j}(N)+iN_{i-1,j}(N).
\end{equation}
Now the substitution
\begin{equation}
\label{Aij-def}
N_{i,j}(N)=\frac{\Gamma(N-i-1)\,\Gamma(i+1)}{\Gamma(N)}\,A_{i,j}(N)
\end{equation}
reduces (\ref{Nij1}) to the constant-coefficient recursion
\begin{equation*}
A_{i,j}(N+1)=A_{i,j}(N)+A_{i-1,j}(N)
\end{equation*}
that allows us to express $A_{i,j}$ via $A_{i-1,j}$:
\begin{equation}
\label{Aij}
A_{i,j}(N+1)=\sum_{M=1}^N A_{i-1,j}(M).
\end{equation}

{}From (\ref{Aij-def}) we find $A_{0,j}(N+1)=N^{-1}N_{0,j}(N+1)$ which, in
conjunction with (\ref{N0j}), gives
\begin{equation}
\label{A0j}
A_{0,j}(N+1)=\sum_{M=1}^NQ_{j-1}(M)
\end{equation}
Therefore starting with $A_{0,j}$ from Eq.~(\ref{A0j}), we find all the
$A_{i,j}$ via Eq.~(\ref{Aij}).  The final result is
\begin{equation}
\label{Aij-sol}
A_{i,j}(N)=\sum_{1\leq m_0<\ldots<m_i<N} Q_{j-1}(m_0)
\end{equation}
Equations (\ref{Qj-sol}), (\ref{Aij-def}), (\ref{Aij-sol}) give the full solution
for the joint degree distribution.

While the complete solution is cumbersome, it can be simplified as
$N\to\infty$.  In this limit, we can first replace the factor
$\frac{\Gamma(N-i-1)}{\Gamma(N)}$ by $N^{-i-1}$ in Eq.~(\ref{Aij-def}).
Additionally, we can replace the summation in Eq.~(\ref{Aij-sol}) by
integration.  These two replacements are justified when $i\ll \sqrt{N}$.
Finally, using (\ref{Qj-sol-cont}) and after some algebra, we find the
leading behavior
\begin{equation*}
N_{0,j}(N)\to \frac{(\ln N)^{j-1}}{(j-1)!}
\end{equation*}
and more generally
\begin{equation}
\label{Nij-sol-cont}
N_{i,j}(N)\to \left[\frac{(\ln N)^{j-1}}{(j-1)!}\right]^{i+1}.
\end{equation}
Because the Poisson form the out-degree distribution holds only when $j\ll
\ln N$, the generalized Poisson form (\ref{Qj-sol-cont}) for the joint degree
distribution is also valid only for $j\ll \ln N$.

Finally, although the total degree distribution $N_k(N)$ does not satisfy a
closed equation, we can obtain this distribution indirectly.  When $k$ is of
the order of $\ln N$ or smaller, we can use (\ref{Nij-sol-cont}) to find
$N_k(N)$.  The situation in the range $k\gg \ln N$ is even simpler: In this
region, the total degree distribution essentially coincides with the
in-degree distribution and therefore $N_k(N)\to N\,k^{-2}$ (Fig.~\ref{deg-dist}).

\section{Summary}

We introduced a growing network model that is based on node addition plus a
simple copying mechanism --- the GNC --- that leads to an average node degree
growing logarithmically with the total number of nodes $N$.  This feature may
account for the intriguing phenomenon observed in many real networks that the
number of links increases slightly faster than the number of nodes.  Copying
arises naturally in the context of citations; a not untypical scenario is
that an author will be familiar with a few primary references, but may simply
copy secondary references from primary ones.

We solved the underlying master equations for the GNC model and showed that
the in-degree distribution is a power-law over its entire range, while the
out-degree distribution is asymptotically Poissonian.  The total degree
distribution is consequently a hybrid of the power-law and Poisson forms.
There is, on average, one node with total degree equal to 1, and there is
always one node --- the root --- that has in-degree equal to $N-1$.  Thus the
node degree ranges from one to $N-1$.  Since the distribution of $L$ has a
width that scales linearly with $N$ while $L(N)$ grows as $N\ln N$,
fluctuations in node degree are appreciable even for very large networks.
Finally, each node is connected to the root, so that the network diameter
equals 2, independent of $N$.

From long-term Physical Review publication data, the average number of
references per paper (the out degree) grows slowly with the total literature
size, consistent with the logarithmic growth predicted by the GNC model.
However, this growth in the GNC model is not robust when parameters that
quantify the extent of copying are varied.  The apparent logarithmic growth
for the average number of references per paper in the Physical Review is thus
a bit surprising and it will be worthwhile to test whether logarithmic growth
arises in a wider range of empirical networks.

\acknowledgments{We thank Tibor Antal and Matt Hastings for helpful comments,
  as well as Mark Newman for informing us about Ref.~\cite{ENP}.  One of us
  (SR) is grateful for financial support by NSF grant DMR0227670 (at BU) and
  DOE grant W-7405-ENG-36 (at LANL). }

\appendix
\section{Fluctuations} 
\label{ap}

In this appendix, we find the variance in the distribution of the number of
links $P_N(L)$.  We start by computing the first two moments of the
out-degree distribution, $u_N\equiv\langle j\rangle=\sum_j j Q_j(N)/N$ and
$v_N\equiv\langle j^2\rangle$.  We then use these results to derive the
variance of $P_N(L)$.

To determine $u_N$ and $v_N$, we can in principle use Eq.~(\ref{Qj-sol}).
However, a direct approach is more useful.  Starting with $Nu_N=\langle \sum
j\rangle$, we find that adding a new node leads to the recursion relation
\begin{eqnarray}
\label{uN}
(N+1)u_{N+1}&=&\left\langle 1+j_\alpha+\sum j\right\rangle\nonumber\\
           &=&1+u_{N}+Nu_{N},
\end{eqnarray}
which is nothing but Eq.~(\ref{LN}).  In a similar manner, we derive a
recursion relation for $Nv_N=\langle \sum j^2\rangle$
\begin{eqnarray*}
(N+1)v_{N+1}&=&\left\langle (1+j_\alpha)^2+\sum j^2\right\rangle\\
           &=&1+2u_{N}+v_N+Nv_{N},
\end{eqnarray*}
which reduces to 
\begin{equation}
\label{vN}
v_{N+1}=v_{N}+\frac{2}{N+1}\,u_N+\frac{1}{N+1}.
\end{equation}
The variance $\sigma^2(N)=v_N-(u_N)^2$ therefore satisfies 
\begin{equation}
\label{sN}
\sigma^2(N+1)=\sigma^2(N)+\frac{1}{N+1}-\frac{1}{(N+1)^2}.
\end{equation}
{}From this simple recursion we get 
\begin{equation}
\label{sN-sol}
\sigma^2(N)=H_N-H_N^{(2)}.
\end{equation}
The relative magnitude of fluctuations die out slowly, as the standard
deviation, $\sigma(N)\sim \sqrt{\ln N}$ divided by the average $u_N\sim \ln N$,
approaches zero as $(\ln N)^{-1/2}$.

Consider now the average number of links in the network $L(N)=\langle \sum
j\rangle=Nu_N$ and the corresponding second moment $L_2(N)\equiv\langle
L_N^2\rangle$.  After the addition of a new node, the second moment changes
according to
\begin{eqnarray*}
L_2(N\!+\!1)\!\!&=&\!\!\left\langle \left(1+j_\alpha+\sum
                       j\right)^2\right\rangle\\ 
&=&\!\!\left\langle (1\!+\!j_\alpha)^2\!+\!2\sum j \!+\!2j_\alpha\sum j 
\!+\!\left(\sum j\right)^2\right\rangle\\
   &=&1\!+\!2u_{N}\!+\!v_N\!+\!2Nu_{N}\!+\!\left(1\!+\!\frac{2}{N}\right) L_2(N)
\end{eqnarray*}
Now we use $L(N)=\langle\sum j\rangle =Nu_N$ to write the square of
Eq.~(\ref{uN}) in the form
\begin{equation*}
L(N+1)^2=\left(1+\frac{2}{N}\right)L(N)^2+(u_N)^2+2(N+1)u_{N}+1,
\end{equation*}
and then subtracting this from the previous equation, the variance
$\Sigma^2(N)= L_2(N)-L(N)^2$ satisfies
\begin{equation}
\label{SN}
\Sigma^2(N+1)=\left(1+\frac{2}{N}\right)\Sigma^2(N)+\sigma^2(N)
\end{equation}

The homogeneous part of (\ref{SN}) suggests seeking a solution of the form
$\Sigma^2(N)=N(N+1)S_N$.  This substitution recasts (\ref{SN}) into
$S_{N+1}-S_N=[(N+1)(N+2)]^{-1}\sigma^2(N)$.  This is an exact discrete first
derivative for $S_N$.  Hence $S_N$ equals
$\sum_1^{N-1}[(M+1)(M+2)]^{-1}\sigma^2(M)$.  Thus the variance is
\begin{equation}
\label{SN-sol}
\Sigma^2(N)=N(N+1)\sum_{M=1}^{N-1} \frac{\sigma^2(M)}{(M+1)(M+2)}.
\end{equation}
Finally, by substituting $\sigma^2(M)=\sum_{j\leq M}(j^{-1}-j^{-2})$ from
(\ref{sN-sol}) into (\ref{SN-sol}), and changing the order of the two sums,
we find that $\Sigma^2(N)\to (2-\frac{1}{6}\,\pi^2)N(N+1)$ as $N\to\infty$.
This leads to the asymptotic expression for the standard deviation given in
Eq.~(\ref{SN-sol-asymp}).


\begin{thebibliography}{99}
  
\bibitem{rev} R.~Albert and A.-L.~Barab{\'a}si, Rev.\ Mod.\ Phys.\ {\bf 74},
  47 (2002); M.~E.~J.~Newman, SIAM Review {\bf 45}, 167 (2003).

\bibitem{book} S.~N.~Dorogovtsev and J.~F.~F. Mendes, {\em Evolution of
    Networks: From Biological Nets to the Internet and WWW} (Oxford
  University Press, Oxford, 2003); R.~Pastor-Satorras and A.~Vespignani, {\em
    Evolution and Structure of the Internet: A Statistical Physics Approach}
  (Cambridge University Press, Cambridge, 2004).

\bibitem{alex} 
  A.~Vazquez, Europhys.\ Lett.\ {\bf 54}, 430 (2001); A.~Vazquez, cond-mat/0105031.

\bibitem{grandson} 
  G. Yan, T. Zhou, Y.-D. Jin, and Z.-Q. Fu, cond-mat/0408631.

\bibitem{DM} S.~N.~Dorogovtsev and J.~F.~F.~Mendes, in {\it Handbook of
    Graphs and Networks: From the Genome to the Internet}, eds.\ S. Bornholdt
  and H.G. Schuster (Wiley-VCH, Berlin, 2002). 

\bibitem{sen} P. Sen, Phys.\ Rev.\ E {\bf 69}, 046107 (2004);
  cond-mat/0409154.

\bibitem{KR01} P.~L.~Krapivsky and S.~Redner, Phys.\ Rev.\ E {\bf 63}, 066123
  (2001).

\bibitem{KR021} P.~L.~Krapivsky and S.~Redner, J. Phys.\ A {\bf 35}, 9517
  (2002); P.~L.~Krapivsky and S.~Redner, Phys.\ Rev.\ Lett.\ {\bf 89}, 258703
  (2002).

\bibitem{kum} J.~Kleinberg, R.~Kumar, P.~Raghavan, S.~Rajagopalan, and
  A.~Tomkins, in: {\it Proceedings of the International Conference on
    Combinatorics and Computing}, Lecture Notes in Computer Science,
  Vol.~1627, pp.\ 1-18 (Springer-Verlag, Berlin, 1999).
   
\bibitem{sim} M.~V.~Simkin and V.~P.~Roychowdhury, Complex Syst.\ {\bf 14},
  269--274 (2003).

\bibitem{knuth} R.~L.~Graham, D.~E.~Knuth, and O.~Patashnik, {\it Concrete
    Mathematics\,: A Foundation for Computer Science} (Reading, Mass.:
  Addison-Wesley, 1989).
   
\bibitem{note} In writing Eq.~(\ref{LN-mpq}) we ignore that for $m\geq 2$ (i)
  the same node can be chosen twice as the target, and (ii) the same node can
  be chosen as the ancestor of target nodes. As long as $mq<2$ our results
  are asymptotically exact.

\bibitem{PR} S. Redner, physics/0407137.
  
\bibitem{new} M. E. J. Newman, Phys.\ Rev.\ E {\bf 64}, 016131 (2001); Phys.\ 
  Rev.\ E {\bf 64} (2001) 016132; Proc.\ Natl.\ Acad.\ Sci.\ USA {\bf 98},
  404 (2001); Proc.\ Natl.\ Acad.\ Sci.\ USA {\bf 101}, 5200 (2004).

\bibitem{dor} J.~J.~Ramasco, S.~N.~Dorogovtsev, and R.~Pastor-Satorras,
  Phys.\ Rev.\ E {\bf 70}, 036106 (2004).

\bibitem{LJL} S. Lehmann, A. D. Jackson, and B. Lautrup, {\it cond-mat/0408472}.

\bibitem{ENP} http://www.oakland.edu/enp/collab.pdf.  This publication is
  part of the website devoted to the Erd\H{o}s Number project.  See
  http://www.oakland.edu/enp.

\bibitem{broder} A.~Broder, R.~Kumar, F.~Maghoul, P.~Raghavan,
  S.~Rajagopalan, R.~Stata, A.~Tomkins, and J.~Wiener, Computer Networks {\bf
    33}, 309--320 (2000).

\bibitem{don} D.~Donato, L.~Laura, S.~Leonardi, and S.~Millozzi, Eur.\ Phys.\
  J. B {\bf 38}, 239--243 (2004).
   
\bibitem{BNC} A. Broido, E. Nemeth, and kc claffy, Eur.\ Trans.\ on
  Telecommunications, January 2002.

\bibitem{AH} 
  B.~A.~Huberman and L.~A.~Adamic, Nature {\bf 401}, 131 (1999).

\bibitem{BA} 
  A.-L.~Barab\'asi and R.~Albert, Science {\bf 286}, 509 (1999).

\bibitem{www} 
  S.~R.~Kumar, P.~Raghavan, S.~Rajagopalan, and A.~Tomkins, in:
  {\it Proc. 8th WWW Conf.} (1999).
   
\bibitem{stanford} The Stanford WebBase project:
  http://www-diglib.stanford.edu/tested/doc2/WebBase/


\end{thebibliography}
\end{document}